\documentclass[12pt,preprint]{aastex}

\def\kms{\ifmmode{\rm km\thinspace s^{-1}}\else km\thinspace s$^{-1}$\fi}

\shortauthors{Torres et al.}
\shorttitle{TW~Hya association}

\begin{document}

\title{Radial-velocity survey of members and candidate members of the
TW~Hydrae association\altaffilmark{1}}

\altaffiltext{1}{Some of the observations reported here were obtained 
with the Multiple Mirror Telescope, a joint facility of the
Smithsonian Institution and the University of Arizona.}

\email{** To appear in the February 2003 issue of The Astronomical Journal **}

\author{Guillermo Torres\altaffilmark{2},
	Eike W.\ Guenther\altaffilmark{3},
	Laurence A.\ Marschall\altaffilmark{4},
	Ralph Neuh\"auser\altaffilmark{5},
	David W.\ Latham\altaffilmark{2},
	Robert P.\ Stefanik\altaffilmark{2}
}

\altaffiltext{2}{Harvard-Smithsonian Center for Astrophysics, 60
Garden St., Cambridge, MA 02138}

\altaffiltext{3}{Th\"uringer Landessternwarte Tautenburg,
Karl-Schwarzschild-Observatorium, Sternwarte 5, 07778 Tautenburg,
Germany}

\altaffiltext{4}{Department of Physics, Gettysburg College, 300 North
Washington Street, Gettysburg, PA 17325}

\altaffiltext{5}{Max-Planck-Institut f\"ur extraterrestrische Physik,
D-85740 Garching, Germany}

\email{gtorres@cfa.harvard.edu}

\begin{abstract}

We report our spectroscopic observations of stars belonging to the
young nearby group known as the TW~Hydrae association, as well as of a
number of potential members of the association identified in kinematic
and X-ray surveys. Multiple radial velocity measurements were obtained
for each object, several of which turn out to be multiple systems.
Orbital solutions are presented for 3 double-lined binaries, one
single-lined binary, and a double-lined triple system, all with short
periods. Effective temperatures and projected rotational velocities
are presented for each visible object. None of the candidate members
of the association in our sample is confirmed as a true member. The
large fraction of close binaries among the candidate members has to do
with their selection based on X-ray emission from ROSAT, which tends
to favor the inclusion of tidally-locked systems that are active but
not necessarily young. 
	
\end{abstract}


~\newpage

\section{Introduction} \label{sec:introduction} In recent years a
number of loose associations of nearby stars have been identified that
appear to be very young, yet show no signs of molecular gas in the
surroundings ---one of the more visible characteristics of the
classical regions of star formation. Among these new groups are the
TW~Hydrae association \citep{Kastner1997}, the $\eta$~Chamaeleontis
cluster \citep{Mamajek1999}, the Tucana association
\citep{Zuckerman2000}, the Horologium association \citep{Torres2000}
\citep[which may be the same or related to the latter;][]
{Zuckerman2001,Guenther2001}, the Capricornius association or
HD~199143 group \citep{vandenAncker2001}, which is probably related or
is a subgroup of the $\beta$~Pictoris group \citep{Zuckerman2001a},
and others. Other small groups are similarly young and nearby but do
appear to be associated with gas, such as the stars in the
high-latitude MBM12 cloud \citep{Hearty2000a,Hearty2000b}. The
distances to these groups are in the range from $\sim$35 to
$\sim$150~pc, and they all seem to have similar ages around 10~Myr.
The fact that most of them are in the southern sky is probably not a
coincidence, and may be related to their association with the Gould
Belt, or more specifically, the Sco-Cen association and its subgroups
\citep[see][]{Mamajek2000,Mamajek2001}. 

The best known of these small groups of young stars is the TW~Hya
association, named after the first classical T~Tauri star found in
isolation from any known cloud material
\citep{Henize1976,Herbig1978,Rucinski1983}. Surveys by
\cite{delaReza1989} and \cite{Gregorio1992} based on the IRAS
Point-Source Catalog turned up four other T~Tauri stars in the same
region of the sky, and the physical association between all these
stars was suggested by \cite{Kastner1997} on the basis of the
similarity of their optical as well as their X-ray properties.
Subsequent systematic searches revealed other apparently related
stars, after it was found that they also appear to share a similar
space velocity. A few previously known young stars in the same area of
the sky with similar characteristics were also included in the group.
To date there are some 20 recognized members spread over hundreds of
square degrees
\citep{Kastner1997,Webb1999,Sterzik1999,Zuckerman2001b}, along with a
number of other candidate members identified on the basis of their
kinematics, their X-ray properties, or their infrared (2MASS) colors
and spectral features \citep[see, e.g.,][]{Gizis2002}. The typical
distance of these objects is roughly 60~pc, although there appears to
be a significant spread. 

Many of them have been the subject of a broad range of studies to
characterize their properties and to establish their youth, by
measuring the strength of their H$\alpha$ emission,
\ion{Li}{1}~$\lambda$6708 absorption, infrared excess, etc. Other
high-resolution imaging investigations have focussed on circumstellar
disks (TW~Hya, HR~4796A, Hen~3-600A, HD~98800B) and binary companions,
and have even revealed the presence of a probable brown dwarf around
one of the stars
\citep[CD$-33\arcdeg$7795;][]{Lowrance1999,Webb1999,Neuhauser2000a}. 

Kinematic investigations relying on the assumption of a common space
motion for the members of the TW~Hya association (convergent point
solution) have been carried out by \cite{Makarov2001} and also
\cite{Frink2001} to study the structure of the group, using the proper
motions of the known members and trigonometric parallaxes from the
HIPPARCOS mission for the few stars that have them. In this way
additional members have been proposed, and radial velocities have been
predicted for the known and candidate stars by \cite{Makarov2001}.
Direct measurements of the radial velocities of these stars have been
made rather sparingly over the past few years by a number of authors,
and occasional discrepancies have shown up. Observations for most of
these new candidates have been reported recently by \cite{Song2002}.
Essentially all of these studies are based on a single measurement of
the velocity of each star or on observations over a very limited time
interval, rather than on a systematic monitoring over time. Since some
of the objects may be binaries, this could explain some of the
differences mentioned above. 

In this paper we present the results of our radial-velocity monitoring
of members and candidate members of the TW~Hya association over the
past several years, with multiple observations per object.  This has
allowed us to solve for the spectroscopic orbits of several binaries
as well as one triple system, and to determine the physical properties
(effective temperature, projected rotational velocity) of all visible
components.  A preliminary report of this work was given by
\cite{Torres2001}, and the full details based on additional
observations are given here. 
	
\section{Sample and observations}
\label{sec:observations}

Potential members of the TW~Hya association were drawn from reports by
\cite{Hoff1997}, \cite{Hoff1998}, \cite{Sterzik1999}, and
\cite{Makarov2001}, and a few other objects were added on the basis of
their proximity to the known members and the similarity of their X-ray
properties from ROSAT. In addition, 7 known members were also
observed, including their binary companions when possible.
Table~\ref{tab:sample} lists the optical properties of the stars in
our sample.  Conventional designations in the association (e.g.,
TWA-1) are given in column (2), and other columns present alternate
designations, coordinates, visual magnitudes, spectral types, and
\ion{Li}{1}~$\lambda$6708 equivalent widths.  We have included in this
list the quadruple system HD~98800 (TWA-4) for completeness, given
that it was observed several years ago with the same instrumental
setup described below \citep{Torres1995}. The X-ray properties of
these stars are given in Table~\ref{tab:xray}. They include ROSAT
positions and their uncertainty (close binaries are unresolved), the
Maximum Likelihood estimator $ML$ that provides a measure of the
existence of the source above the local background
\citep{Cruddace1988}, the exposure time and count rate, and the X-ray
hardness ratios \citep[see][]{Neuhauser1995}. These data are taken from
the ROSAT All-Sky Survey Catalog 1RXS \citep{Voges1999}\footnote{The
catalog and updates are available at {\tt
http://www.xray.mpe.mpg.de/rosat/survey/rass-bsc/}.}.  The majority of
the spectroscopic observations for the present investigation were
obtained with various telescopes at the Harvard-Smithsonian Center for
Astrophysics (CfA), and a few also at the 1.5-m ESO telescope at La
Silla (Chile) and the 2-m telescope in Tautenburg (Germany). 
	


Observations at the CfA were made using nearly identical echelle
spectrographs on the 1.5-m Wyeth reflector at the Oak Ridge
Observatory (Harvard, Massachusetts), the 1.5-m Tillinghast reflector
at the F.\ L.\ Whipple Observatory (Mt.\ Hopkins, Arizona) and the
Multiple Mirror Telescope (also on Mt.\ Hopkins, Arizona) prior to its
conversion to a monolithic 6.5-m mirror. A single echelle order was
recorded with photon-counting intensified Reticon detectors at a
central wavelength of 5187~\AA, with a spectral coverage of 45~\AA.
The resolving power is $\lambda/\Delta\lambda\approx 35,\!000$, and
the signal-to-noise (S/N) ratios achieved range from about 7 to 50 per
resolution element of 8.5~\kms. A total of 509 spectra were collected
over a period of 18 years (1984--2002), including archival
observations of similar quality for some of the objects that were
observed with the same instruments prior to the start of this project. 

Radial velocities were obtained using the cross-correlation task {\tt
XCSAO} \citep{Kurtz1998} running under IRAF\footnote{IRAF is
distributed by the National Optical Astronomy Observatories, which is
operated by the Association of Universities for Research in Astronomy,
Inc., under contract with the National Science Foundation.}.  Typical
uncertainties for an individual measurement are smaller than 1~\kms.
For stars with temperatures hotter than $\sim$4000~K we used templates
from a grid of synthetic spectra computed for us by Jon Morse, based
on the latest model atmospheres by R.\ L.\ Kurucz\footnote{Available
at {\tt http://cfaku5.harvard.edu}.} \citep[see][]{Nordstrom1994}.
These calculated spectra are available for a wide range of effective
temperatures ($T_{\rm eff}$), projected rotational velocities ($v \sin
i$), surface gravities ($\log g$) and metallicities. The optimum
template for each object was determined from extensive grids of
correlations in temperature and rotational velocity (the two
parameters that affect the radial velocities the most), for an adopted
surface gravity and for solar metallicity.  We adopted for the stellar
properties the parameters giving the highest correlation averaged over
all exposures, interpolated between neighboring templates for higher
accuracy.  The errors for the effective temperature and $v \sin i$
determinations are estimated to be around 150~K and 2--3~\kms,
respectively, unless noted otherwise.  For objects cooler than about
4000~K the synthetic templates become less realistic because they lack
several key molecular opacity sources. In those cases we used observed
templates from strong exposures of late-type stars. 

Several of our objects turned out to have composite spectra (two sets
of lines present).  For those we determined the radial velocities
using TODCOR \citep{Zucker1994}, which is a two-dimensional
cross-correlation technique well suited to our relatively low S/N
observations.  Grids of correlations analogous to those described
above were run to determine the stellar properties for both
components, whenever possible. The temperature and $v \sin i$
determinations for all visible objects are given in
Table~\ref{tab:teffrot}, and are compared with similar determinations
by other authors. The agreement in most cases is reasonably good. 

	
The stability of the zero-point of the CfA velocity system was
monitored by means of exposures of the dusk and dawn sky, and small
systematic run-to-run corrections were applied in the manner described
by \citet{Latham1992}. The zero point of the native CfA velocity
system based on synthetic templates is very close to the absolute
frame as defined by extensive observations of the minor planets in the
solar system. The correction required to place our radial velocities
on this absolute frame is $+0.139~\kms$
\citep{Stefanik1999,Latham2002}. 

Forty-two additional observations for four of the objects were
obtained with the echelle spectrograph FEROS (Fiber-fed Extended Range
Optical Spectrograph) on the 1.5-m ESO telescope at La Silla. The
wavelength coverage is approximately from 3600~\AA\ to 9200~\AA\ (38
echelle orders), and the resolving power is
$\lambda/\Delta\lambda\approx 48,\!000$. Due to the relative faintness
of the stars the observations were carried out in `object+sky' mode
rather than in the mode in which calibrations are taken simultaneously
with the object.  We obtained 3 spectra of TWA-1 (TW~Hya), 3 of TWA-2A
(CD$-29\arcdeg$8887A), 18 of TWA-3A (Hen~3-600A), and 18 spectra of
TWA-5A (CD$-33\arcdeg$7795A).  The S/N ratios for these observations
range from 40 to 70 per 0.03~\AA\ pixel at $\lambda$6708.  The
standard MIDAS pipeline for FEROS was used to subtract the bias,
flat-field, remove the scattered light, subtract the sky background,
and to extract and wavelength-calibrate the spectra.  Telluric lines
were used to determine the instrumental shift between the observed
spectra and the Th-Ar comparison spectra taken at the beginning and at
the end of each night. Radial velocities were determined by measuring
the position of photospheric lines in the spectra, and then applying
the instrumental shift along with the barycentric correction. Tests
showed that the error of the radial velocities derived in this way is
approximately 0.3--0.4~\kms. 

An additional spectrum of HIP~53486 was taken with the Coud\'e echelle
spectrograph on the 2-m Alfred Jensch telescope in Tautenburg
(Germany). Use of a 1\farcs2 slit yielded a resolution of
$\lambda/\Delta\lambda\approx 67,\!000$, and the wavelength region
covered is 4680--7400~\AA. Standard IRAF routines were used to
flat-field and extract the spectra. Th-Ar comparison lamp spectra at
the beginning and end of the night were used to establish the
wavelength reference, and instrumental shifts were determined using
telluric lines. 

In addition to the radial velocity determinations we also measured the
strength of the \ion{Li}{1}~$\lambda$6708 absorption line from the
FEROS and Tautenburg spectra. This is one of the classical indicators
of stellar youth \citep[see, e.g.,][]{Bodenheimer1965,Skumanich1972}.
Our equivalent width measurements are $0.455 \pm 0.016$~\AA\ for
TWA-1, $0.560 \pm 0.010$~\AA\ for TWA-2A, $0.658 \pm 0.020$~\AA\ for
TWA-5A, and an upper limit of 0.02~\AA\ for HIP~53486. These have been
combined with other measurements from the literature and listed as
averages in Table~\ref{tab:sample}. 

\section{Stars with orbital solutions}
\label{sec:orbits}

Multi-epoch observations for a number of our objects revealed obvious
velocity variations, or double or distorted peaks in the correlation
functions indicating the presence of a companion. Radial velocities
for each component were measured whenever possible. In several cases
we were able to derive spectroscopic orbital solutions, which we
describe here separately for each system. None of these objects turn
out to be true members of the TW~Hya association, with the exception
of the previously known case of TWA-4~A/B (HD~98800~A/B). Two other
stars that are clearly multiple have so far defied all our attempts to
establish the orbits, but will continue to be observed to that end and
will be the subject of a future paper.  They are TWA-3A (Hen~3-600A)
and TWA-5A (CD$-33\arcdeg$7795A), which are bona-fide members of the
association and have been recognized previously by other authors as
being spectroscopic binaries
\citep{Webb1999,Muzerolle2000,Torres2000}. Both cases are complicated
by the presence of visual companions. The secondary of TWA-3A (TWA-3B)
is at a separation of 1\farcs44 and has a magnitude difference of
$\Delta V\sim 0.5$~mag.  The close visual companion of TWA-5A was
recently discovered using adaptive optics at a separation of only
0\farcs06 \citep{Macintosh2001} and a nearly equal brightness as the
primary, and is different from the very faint TWA-5B (separation
$\sim2\arcsec$), and possibly different also from the spectroscopic
companion. A brief discussion of each of our orbital solutions
follows.

\subsection{HIP~48273}

This object (also known as 4~Sex, HD~85217, and HR~3893) was proposed
as a possible member of the TW~Hya association by \cite{Makarov2001},
although they concluded from its kinematics that it is most likely not
a true member.  Double lines revealing the binary nature of the star
were originally discovered by \cite{Shajn1932}, and preliminary
orbital solutions were published by \cite{Popper1948},
\cite{Popper1949}, and \cite{Mayor1987}, with a period of 3.05~d and
an insignificant eccentricity. Systematic trends in the residuals of
the orbit over the 16-yr interval of observation were reported by
\cite{Popper1948}, suggesting the possible presence of another star in
the system.  However, our higher quality solution based on nearly
twice as many observations over a comparable period of time shows no
such trend.  \cite{Mayor1987} reported a significant decrease in the
velocity semiamplitudes $K_A$ and $K_B$ from a comparison between
their orbit and that of \cite{Popper1948}, which is based on
observations obtained some 42~yr earlier. They interpreted this change
as an indication that there is a third star in the system, which
causes a precession of the node of the orbit of the binary that
results in a slow and periodic change in the inclination angle. Our
long coverage with virtually no change in the instrumental setup
allows us to examine this claim by dividing our observations into two
independent data sets, with a difference in the mean epochs of
$\sim$10~yr.  No significant differences are seen in the velocity
amplitudes, despite our uncertainties in the $K$ values being much
smaller than previous solutions. 

\begin{figure}
\epsscale{1.0}
\vskip -1in
\plotone{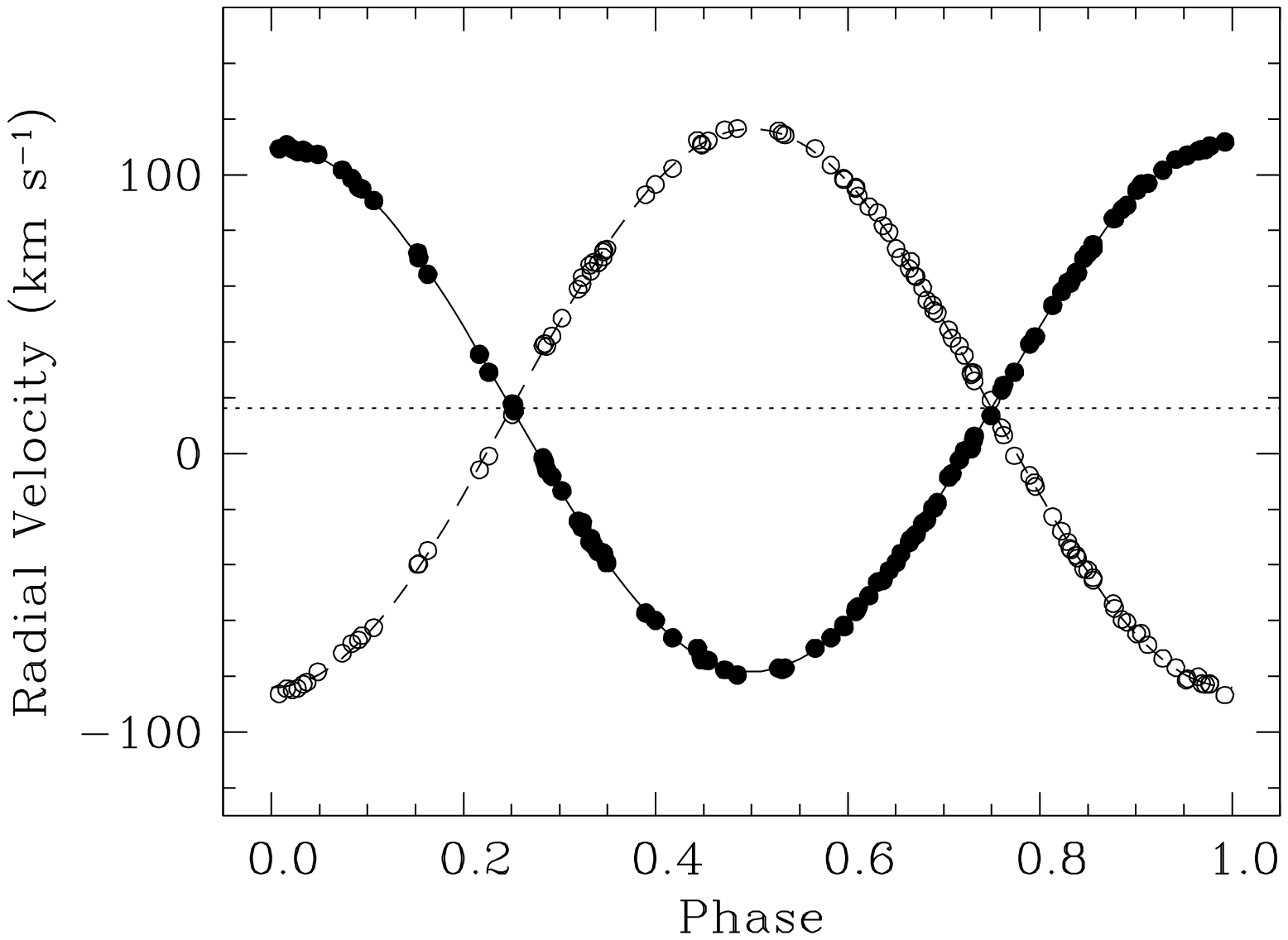}
\vskip -1in
 \caption[Torres.fig1.ps]{Radial velocity observations for
HIP~48273 (filled circles for the primary) along with our orbital
solution. The center-of-mass velocity is indicated by the dotted
line.\label{fig:hip48273}}
 \end{figure}

Table~\ref{tab:hip48273rvs} lists our
radial-velocity measurements, and Table~\ref{tab:hip48273elem} gives
the elements of the orbital solution.  The orbit is circular. The fit
is displayed in Figure~\ref{fig:hip48273}. 


From the short orbital period one may assume that the components'
rotation is synchronized with the orbital motion and their spin axes
are parallel to the axis of the orbit. The system is then detached,
and the minimum inclination angle for eclipses to occur is $i_{\rm
min} \approx 77\arcdeg$ \citep[see][]{Torres2002}. Assuming normal
masses for stars of the temperatures we determine \citep[e.g., based
on the tabulation by][]{Gray1992}, the actual inclination angle must
be close to $i_{\rm min}$. However, no eclipses are seen in the epoch
photometry provided by the HIPPARCOS mission, which displays a scatter
of only 7~millimagnitudes\footnote{Two of the HIPPARCOS observations
show a brightness $\sim$0.15~mag fainter than the average, and happen
to occur at phases 0.282 and 0.726, very near the predicted times of
eclipse at phase 0.25 and 0.75. Although this would appear to suggest
perhaps grazing eclipses, other observations at similar phases show
the object at normal brightness, indicating that the low points are
probably due to measurement errors.}. 

\placefigure{fig:hip48273}

The light ratio between the primary and secondary at the wavelength of
our observations (5187~\AA) is $l_B/l_A = 0.66 \pm 0.02$. The mass
ratio and the light ratio are consistent with the mass-luminosity
relation for main-sequence stars. This, along with the weak
\ion{Li}{1}~$\lambda$6708 absorption (even correcting for duplicity)
indicates the system is probably not in the pre-main sequence stage,
and therefore is unlikely to be a true member of the association. The
center-of-mass velocity of $+16.3$~\kms\ is also quite different from
the typical value of about $+11$~\kms\ for true members (see
\S\ref{sec:discussion}). 


\subsection{TYC~6604-0118-1}

This soft X-ray source from the Einstein Observatory Extended Medium
Sensitivity Survey \citep{Gioia1990,Stocke1991}, referred to there as
1E0956.8$-$2225, was also proposed as a possible member of the TW~Hya
association by \cite{Makarov2001}. It was originally identified as a
possible spectroscopic binary by \cite{Fleming1988}, and is also known
by the names SAO~178272 and BD$-21\arcdeg$2961, among others.  A
preliminary double-lined spectroscopic orbital solution was reported
by \cite{Stefanik1992} with a period of 1.84 days, and also by
\cite{Baker1994} who derived a small eccentricity of $e = 0.016$. Our
improved solution indicates a circular orbit, and the light ratio
between the stars is $l_B/l_A = 0.39 \pm 0.02$. The velocities and
orbital elements are given in Table~\ref{tab:1e09568rvs} and
Table~\ref{tab:1e09568elem}, respectively. 



\begin{figure}
\epsscale{1.0}
\vskip -1in
\plotone{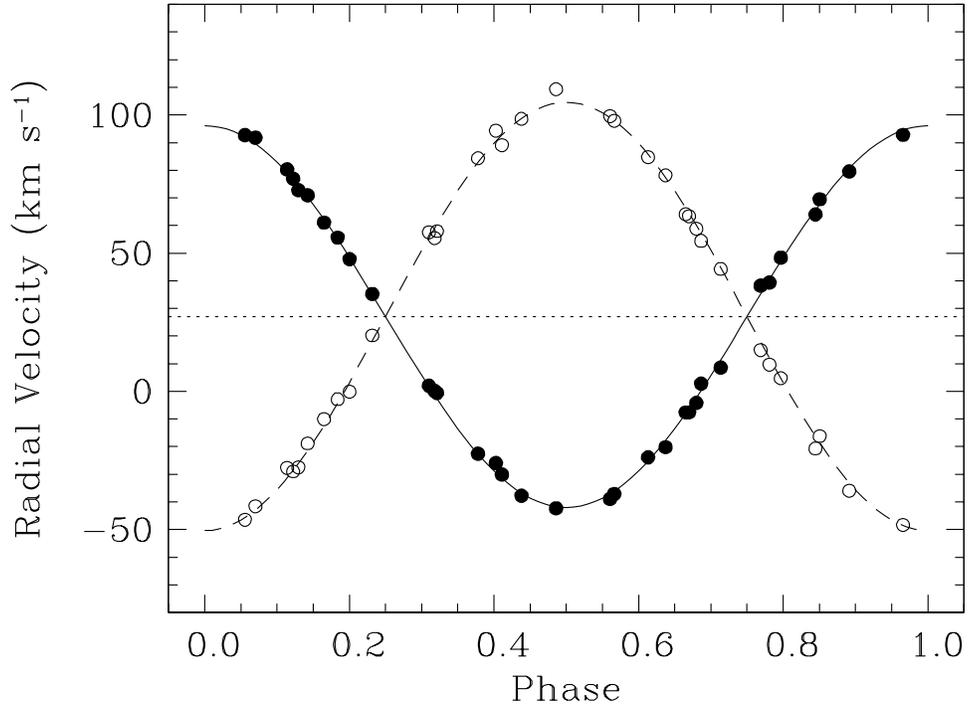}
\vskip -1in
 \caption[Torres.fig2.ps]{Radial velocity observations for
TYC~6604-0118-1 (filled circles for the primary) along with our
orbital solution. The center-of-mass velocity is indicated by the
dotted line.\label{fig:1e09568}}
 \end{figure}

Our orbital fit is shown in Figure~\ref{fig:1e09568}.  As in the
previous case, the light ratio and mass ratio are consistent with the
mass-luminosity relation for dwarfs.  If the measurement of the
\ion{Li}{1}~$\lambda$6708 strength (Table~\ref{tab:sample}) in this
double-lined binary is assumed to correspond to the brighter primary,
a correction for the dilution produced by the light of the secondary
would increase the equivalent width to roughly 0.16~\AA. This is much
weaker than typical Li strengths for stars of similar temperature in
the Pleiades cluster, arguing that TYC~6604-0118-1 is considerably
older than the Pleiades (age $\sim$120~Myr), and hence is not a member
of the TW~Hya association.  Its center-of-mass velocity is also quite
different from that of the member stars. The minimum angle for
eclipses to occur is $i_{\rm min} \approx 77\arcdeg$, whereas the
angle derived assuming normal main-sequence masses for the stars is $i
\approx 48\arcdeg$. The system is detached. 

\placefigure{fig:1e09568}
	
\subsection{RXJ1100.0$-$3813}

This object was selected by us as a possible member based on proximity
on the sky and the X-ray properties from ROSAT. The observations show
it to be a single-lined binary with a period of 1.37~d and a circular
orbit. The measured radial velocities are listed in
Table~\ref{tab:stts-11rvs} and the elements are given in
Table~\ref{tab:stts-11elem}.

\begin{figure}
\epsscale{1.0}
\vskip -1in
\plotone{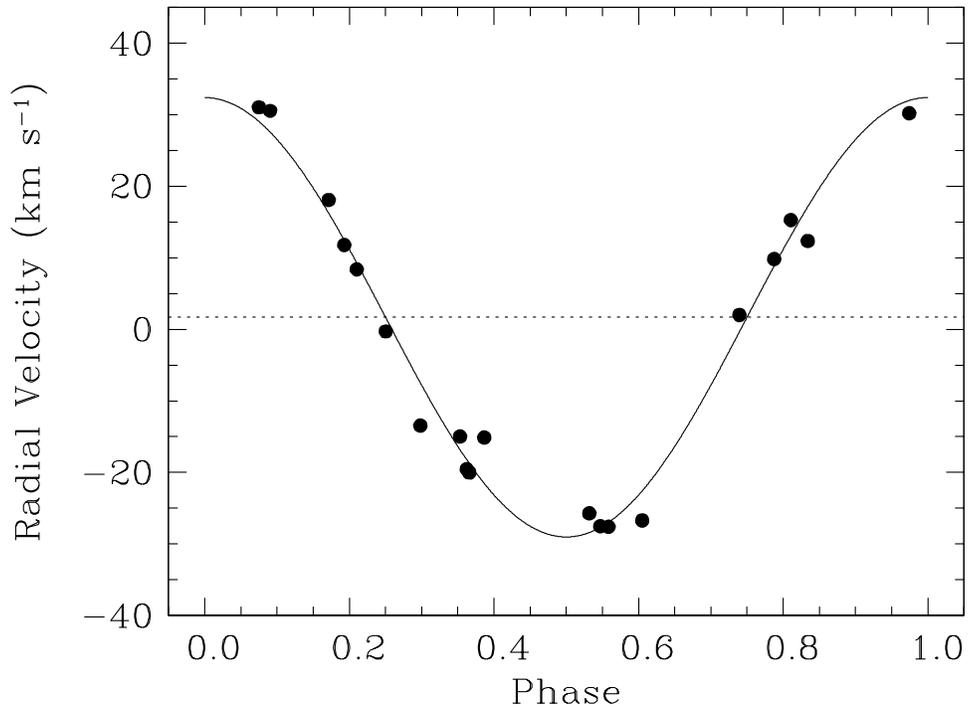}
\vskip -1in
 \caption[Torres.fig3.ps]{Radial velocity observations for
RXJ1100.0$-$3813 along with our orbital solution. The center-of-mass
velocity is indicated by the dotted line.\label{fig:stts-11}}
 \end{figure}

We see no sign of the secondary in our spectra. From an estimated mass
for the primary of 1.0~M$_{\sun}$ based on its derived effective
temperature, we infer a minimum mass for the secondary of
0.18~M$_{\sun}$ if the orbit is viewed edge-on, corresponding to a
spectral type around M5. If the companion is a main sequence star, it
is unlikely to be earlier than about K5 or it would be bright enough
that we would have detected its spectral lines. The orbit and our
measurements are shown in Figure~\ref{fig:stts-11}. On the basis of
its kinematics this object is unlikely to be a true member of the
TW~Hya association. No measurements of the \ion{Li}{1}~$\lambda$6708
strength are available. 
	

\placefigure{fig:stts-11}

\subsection{HD~97131}

Also known as HIP~54610, CD$-29\arcdeg$8898, and RXJ1110.5$-$3027,
this object is another X-ray source from the ROSAT All Sky Survey that
was initially considered a possible member of the TW~Hya association.
The HIPPARCOS mission later showed that it is much more distant than
the other members ($\pi_{\rm HIP} = 4.86 \pm 1.23$~mas), and in
addition the \ion{Li}{1}~$\lambda$6708 line is very weak, both of
which indicate that it is probably not associated. The systemic
velocity is also completely different from that of the known members
in the same area of the sky. 

\begin{figure}
\epsscale{1.05}
\vskip -1.3in
\hskip -0.33in\plotone{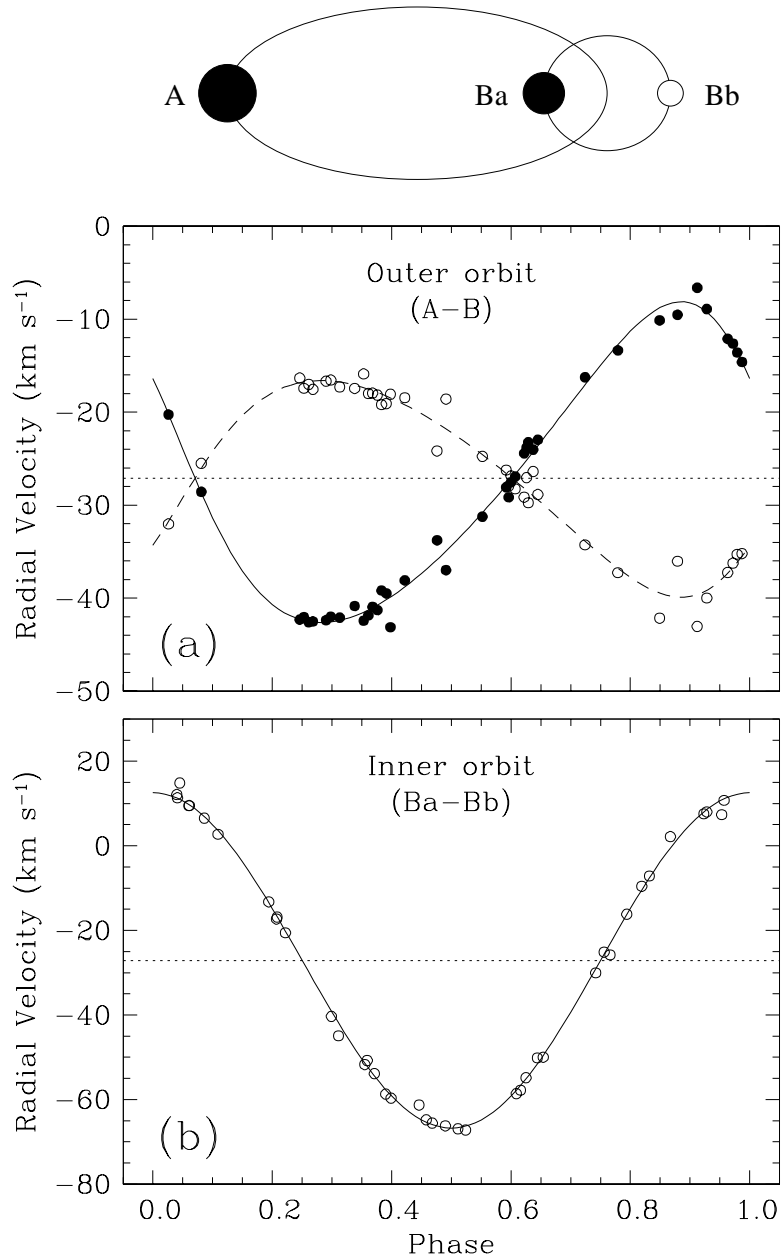}
\vskip -1in
 \caption[Torres.fig4.ps]{Schematic view of the HD~97131 triple
system, and radial-velocity observations along with our orbital
solutions for the inner and outer orbits. The systemic velocity is
indicated by the dotted lines.  (a) Outer orbit with a period of
133~d. The primary (A) is indicated by the filled circles, and the
motion of the secondary (Ba, open circles) in the inner orbit has been
removed; (b) Inner orbit for the secondary, with the velocity of the
visible star corrected for motion in the wide
orbit.\label{fig:stts-8}}
 \end{figure}

The object turned out to be a double-lined triple system. The orbital
period of the primary (the brighter star, ``A") is 133.5~d, and the
secondary is itself a single-lined binary with a period of 1.76~d. The
velocities for the two visible objects (``A" and ``Ba") are given in
Table~\ref{tab:stts-8rvs}. The two orbits were solved simultaneously
under the usual approximation that the inner binary acts as a point
source located at its center of mass for computing the motion in the
outer orbit. Light travel time corrections are relatively small ($<
0.001$~d) but were applied nevertheless, and are given in
Table~\ref{tab:stts-8rvs}.  Table~\ref{tab:stts-8elem} lists the
orbital elements of this combined solution. The orbit of the secondary
is circular, as expected from its short period, while the wide orbit
is slightly eccentric. The fits are shown in Figure~\ref{fig:stts-8},
which includes a schematic view of the system. The light ratio between
the primary and secondary is $l_{\rm Ba}/l_{\rm A} = 0.51 \pm 0.02$. 



Adopting a typical mass for a main-sequence star with the effective
temperature we determine for the primary (which corresponds to SpT
$\sim$ F0), the inclination angle of the wide orbit is estimated to be
$i_{\rm AB} \approx $26$\arcdeg$.  The total mass of the secondary
sub-system implied by this angle is $M \approx 2.3$~M$_{\sun}$.
Adopting a typical mass for star Ba based on its temperature, and
using the mass function of the secondary from our orbital solution, we
can also estimate the inclination angle of the orbit of the secondary
($i_{\rm B}$), which remarkably turns out to be also approximately
26$\arcdeg$.  The inner and outer orbits may thus be
coplanar\footnote{The relative inclination of the two orbits, $i_{\rm
rel}$, depends also on the difference between the position angles of
the ascending nodes ($\Omega$) \citep[e.g.,][]{Fekel1981}, which can
only be determined from astrometry: $\cos i_{\rm rel} = \cos i_{\rm
AB} \cos i_{\rm B} + \sin i_{\rm AB} \sin i_{\rm B} \cos(\Omega_{\rm
AB} - \Omega_{\rm B})$.  Nevertheless, the similarity between $i_{\rm
AB}$ and $i_{\rm B}$ is highly suggestive.}. With these parameters,
the measured $v \sin i$ of star Ba and the assumption of spin-orbit
synchronization lead to a radius for that star consistent with the
expected value for its mass.  The unseen star (Bb) is inferred to have
a mass $M \approx 0.9$~M$_{\sun}$ (SpT $\sim$ G7-G8). 

\placefigure{fig:stts-8}
	
\subsection{RXJ1115.1$-$3233}

\begin{figure}
\epsscale{1.0}
\vskip -1in
\plotone{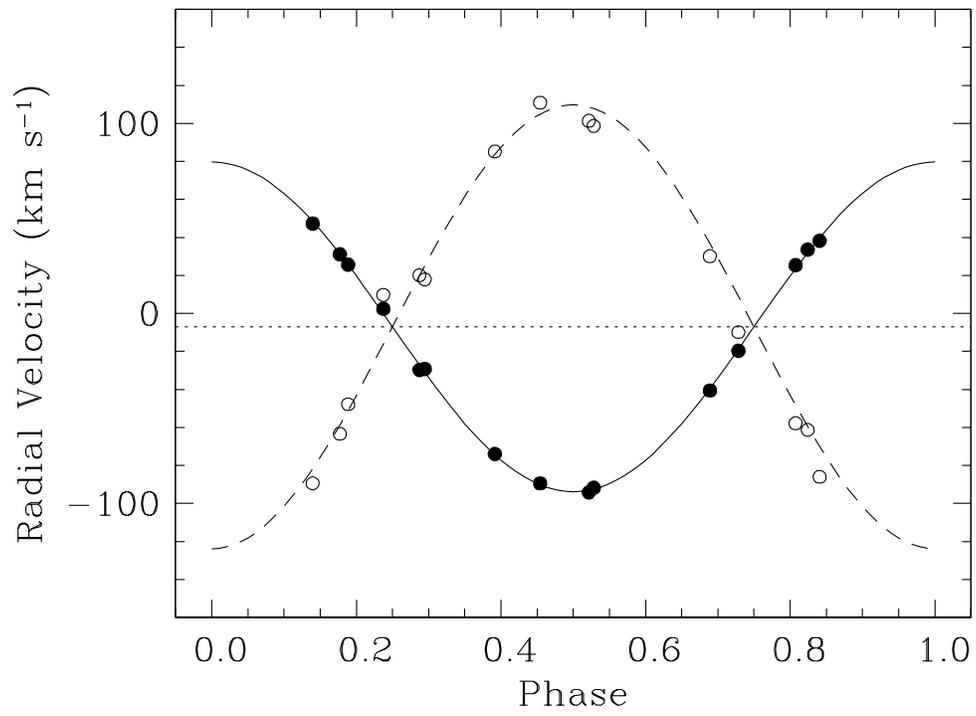}
\vskip -1in
 \caption[Torres.fig5.ps]{Radial velocity observations for
RXJ1115.1$-$3233 (filled circles for the primary) along with our
orbital solution. The center-of-mass velocity is indicated by the
dotted line.\label{fig:stts-10}}
 \end{figure}

As in the previous case, this star was selected as a candidate member
based on its proximity on the sky and its X-ray properties. It is a
double-lined binary with a circular orbit and a period of 2.23~d. The
secondary is quite faint, with the ratio of the brightness of the two
stars being $l_B/l_A = 0.14 \pm 0.02$.  The radial-velocity
measurements are listed in Table~\ref{tab:stts-10rvs}, and the orbital
elements are given in Table~\ref{tab:stts-10elem}. The observations
and the orbit are displayed in Figure~\ref{fig:stts-10}. 



Under the assumption that the stars are rotating synchronously with
the orbit, the minimum angle for eclipses to occur is $i_{\rm min}
\approx 80\arcdeg$. However, the minimum masses are large enough (for
the temperatures we determine) that the possibility of eclipses cannot
be completely ruled out. The system is detached. The mass ratio and
light ratio conform to the typical mass-luminosity relation for the
main sequence, suggesting the components are probably dwarfs. On the
basis of its systemic velocity the system is unlikely to be a true
member of the TW~Hya association. No measurements of the
\ion{Li}{1}~$\lambda$6708 strength are available. 

\placefigure{fig:stts-10}
	
\section{Stars without orbital solutions}
\label{sec:noorbits}

The individual radial velocities for the remaining stars in our sample
are collected in Table~\ref{tab:tabrvs}, with the exception of TWA-3A
and TWA-5A (see above). The correlation peaks for these two objects
display obvious distortions that appear to change considerably on
short time scales (a few days), and are presumably the result of
blends from the lines of two or more stars of similar brightness. A
better understanding of these systems is required before meaningful
velocities can be extracted from our spectra. 


Several other stars on our target list appear to exhibit variations in
their radial velocities that may indicate the presence of companions.
Among these, RXJ1109.7$-$3907 and TWA-12 have one discrepant velocity
each.  TWA-9A shows a scatter among the 18 velocities available that
is about a factor of 2 larger than expected from the internal errors,
but no coherent periodicity is seen. Velocity ``jitter" at this level
could be caused by the presence of spots, which are not uncommon in
young objects such as this. The star was also observed by the
HIPPARCOS mission (HIP~57589), and the photometric measurements show a
dispersion of about 0.07~mag, again roughly twice as large as the mean
internal errors and perhaps consistent with the spot hypothesis. No
periodicity is detected in the HIPPARCOS photometry. 

Our 4 archival spectra for HIP~50796 obtained in 1986 indicate a
velocity drift of more than 4~\kms\ over an interval of 147 days,
which is highly significant compared to the measurement errors.  Six
observations taken 16~yrs later are some 30~\kms\ higher (and show
further changes over a 1-month period), confirming that the object is
a binary (see Table~\ref{tab:tabrvs}). A recent measurement published
by \cite{Song2002} gives an intermediate velocity.  Astrometric
evidence that the star is a binary is also seen in the HIPPARCOS
observations, where acceleration terms (linear changes in the proper
motion components) were found to be significant.  The period of the
system is as yet unknown. 

\section{Discussion}
\label{sec:discussion}

Table~\ref{tab:meanrvs} lists the mean radial velocity for each of our
targets (10 candidate members, and 10 previously known bona-fide
members). For the binaries with solved orbits the value given is the
center-of-mass velocity. A comparison with other values from the
literature, which are typically less accurate, indicates fairly good
agreement. 


The previously recognized members of the TW~Hya group in our sample
(i.e., the ones with TWA- designations in column 2) all have mean
radial velocities within about 3~\kms\ of each other, supporting their
association. In the case of TWA-4 A/B (HD~98800A/B) the 7~\kms\
difference between the components is due to orbital motion of the two
stars around their center of mass, and is quite large because the pair
is currently approaching periastron passage in a fairly eccentric
orbit \citep{Torres1995,Tokovinin1999}. Each of the components is in
turn a spectroscopic binary. Since they are of similar mass
\citep{Soderblom1998}, it may be assumed that the center of mass of
the quadruple system is close to the average of the velocities of the
two visual components, or $+$9.2~\kms.  Similarly, the average of the
velocities of the two components of TWA-13 is $+$12.1~\kms. TWA-2 and
TWA-9 are also binaries, but only the velocities of the primaries have
been measured. 

The lower part of Table~\ref{tab:meanrvs} collects radial velocity
measurements from the literature for other recognized members of the
TW~Hya association that have them. With the possible exception of
TWA-8A (for which no error estimate is available), the velocities for
these stars are again seen to be consistent with those of the other
members. 

The mean radial velocity of the stars currently believed to belong to
the association is thus approximately $+$11~\kms\ in the heliocentric
frame\footnote{The formal average of the velocities of TWA-1, 2, 3, 4,
7, 9, 11, 12, 13, and 19 from Table~\ref{tab:meanrvs} (with binary
components combined into a single value) is $+11.1 \pm 0.4$~\kms; it
is $+10.7 \pm 0.5$~\kms\ if the somewhat lower value for TWA-8A is
included. The uncertainties given correspond to the error of the
arithmetic mean.}.  However, due to their proximity ($\sim$60~pc; but
see below) they are spread across tens of degrees on the sky, and
kinematic studies by \cite{Frink2001} and \cite{Makarov2001} have
shown that because of this it is expected that they will exhibit a
radial velocity gradient across the region, even if they share a
common space velocity. The convergent-point solution by
\cite{Frink2001} also showed that the spread in the ``kinematic"
distance to individual objects within the association (based on the
proper motions and an assumed streaming velocity) is very significant,
ranging from $\sim$30~pc to $\sim$120~pc.  Good agreement was found
between the kinematic distances and the few direct determinations
available from the HIPPARCOS mission. However, it was also found that
the predicted radial velocities for the known members from this model
are too small by several $\kms$. 

The investigation by \cite{Makarov2001} found a similar spread in
distances and radial velocities, but went a step further and
incorporated uniform expansion of the association into the kinematic
model in order to reproduce the observed radial velocities available
at the time.  They carried out a search for new members in an area of
more than 3000 square degrees by selecting objects with X-ray
properties similar to those of known young stars using the ROSAT
Bright Source Catalogue \citep{Voges1999}, and kinematic criteria
based on proper motions for these X-ray sources from the Tycho-2
catalogue \citep{Hog2000}. In this way they proposed 23 new stars as
possible members of the TW~Hya association.  For each of them they
predicted the radial velocity and distance, using their model that
includes an expansion term.  We have observed four of these
candidates, in addition to a number of other bona-fide members also
listed by Makarov \& Fabricius that satisfy their kinematic criteria.
The top section of Table~\ref{tab:makarov} shows these objects, along
with their proper motion and their predicted and measured distances
($D_{\rm kin}$ and $D_{\rm trig}$) and radial velocities ($RV_{\rm
pred}$ and $RV_{\rm obs}$).  The bottom section of the table adds 3
other recognized members of the association with velocity measurements
from other sources (see Table~\ref{tab:meanrvs}). 


The known members typically have measured radial velocities that are
quite close to the predicted values.  TWA-9A, which we discussed
earlier, may be an exception, and we note that its measured distance
also differs from the model prediction (by nearly 4$\sigma$). The
velocity of TWA-19A shows a rather large deviation from the predicted
value as well, although the precision of the observation in this case
is not as good. 

Of the candidate members proposed by \cite{Makarov2001} that we
measured (first 4 entries in Table~\ref{tab:makarov}) HIP~48273 and
TYC~6604-0118-1 have radial velocities that disagree with the
predictions, and HIP~50796 is a binary for which the center-of-mass
velocity is not yet known (see \S\ref{sec:noorbits}) but seems
unlikely to be as low as the expected value of $+13$~\kms.
\cite{Song2002} also reported velocities for these stars, and their
values are consistent with our conclusions.  HIP~53486 is especially
interesting in that it is the nearest candidate member ($\sim$17~pc),
and also the one with the lowest predicted radial velocity.  The
trigonometric parallax as measured by HIPPARCOS agrees very well with
the kinematic distance from the model, and our measured radial
velocity is only 1.8~\kms\ different from the expected value. The
velocity measurement of \cite{Song2002} is consistent once again with
our own determination.  However, the \ion{Li}{1}~$\lambda$6708 line as
measured by those authors is very weak in all 4 of these objects,
which rules them out as young stars and hence as members of the TW~Hya
association. 

In fact, \cite{Song2002} measured most of the other fainter candidate
members proposed by \cite{Makarov2001}, and found that although some
have radial velocities and/or distances similar to the predictions
from the kinematic model, the equivalent width of the Li line is very
small in all but 3 of them.  Those three stars are TYC~7760-0835-1,
TYC~8238-1462-1, and TYC~8234-2856-1, which are among the most distant
candidate members ($D_{\rm pred}$ ranging from 111~pc to 138~pc).
Their measured radial velocities do not agree with the predicted
values particularly well, but are based on a single observation and
could be affected by orbital motion in a binary.  Some evidence for
this was presented by \cite{Makarov2001}, who reported that a
re-reduction of the Tycho-2 observations revealed that both
TYC~7760-0835-1 and TYC~8238-1462-1 appear to be close visual binaries
with angular separations less than 0\farcs5 and magnitude differences
($\Delta V_T$) under 1~mag. Further radial velocity measurements of
these candidates are needed to confirm those indications. 

With the possible exception of these three stars, it seems that no
other new members of the TW~Hya association have emerged from the
kinematic selection by \cite{Makarov2001}. One of the most interesting
results of that analysis was their finding that their model required
some degree of expansion in the association in order to reproduce the
available observations (proper motions, HIPPARCOS distances, and
particularly the radial velocities), at a rate of 0.12~\kms~pc$^{-1}$.
This implies a dynamical age of 8.3~Myr, remarkably close to other
independent estimates that place the age of the association at about
10~Myr.  It has been pointed out, however, that the above kinematic
study included many candidate members that are now believed not to be
true members of the group \citep{Mamajek2001}, as discussed above. At
least 20 such stars out of the 31 objects in the Makarov \& Fabricius
study contribute to determine the location of the convergent point,
the space velocity, the internal velocity dispersion of the
association, and the expansion rate. As more radial velocities become
available for the bona-fide members, it might prove fruitful to refine
the kinematic modeling to see how the inferred properties and the
selection of new candidate members change. 

Optical and kinematical information for the 20 or so objects currently
considered members of the TW~Hya association is scattered throughout
the literature and is sometimes difficult to track down. For the
benefit of the reader we have compiled a list in
Table~\ref{tab:alltwa} that collects accurate coordinates, spectral
types, optical photometry, absolute proper motions, parallaxes (when
available from the Hipparcos mission), mean \ion{Li}{1}~$\lambda$6708
equivalent widths, and mean heliocentric radial velocities for all
objects that have been referred to with a TWA- designation, as well as
for a few of the most likely new additions from recent studies
mentioned earlier. All currenty known visual companions have been
included as well, with the exception of those considered to be
background objects.  A total of at least 11 of the 19 objects are in
visual binary systems or in systems of higher multiplicity (triples
and quadruples) that include also spectroscopic companions. This
represents a fairly high fraction that seems consistent with the large
binary frequency found for young stars in other populations
\citep[see, e.g.,][]{Mathieu1994}.  We point out, however, that
demonstrating that all these stars with TWA- designations are truly
associated with each other is by no means trivial. In fact,
suggestions that some of them may actually belong instead to other
nearby groups have already been made
\citep[e.g.,][]{Mamajek2001,Song2002}, so it is entirely possible that
future kinematic studies may reduce the list of true members. 

	
Referring back to Table~\ref{tab:sample}, it is rather striking that
of the 10 candidate members that we have examined in this paper (none
of which turned out to be a true member of the TW~Hya association), no
less than 6 are confirmed binaries. Four are double-lined and the
other two single-lined. Furthermore, of the binaries with orbital
solutions, \emph{all of them} have periods of 3 days or less. This is
most likely due to a selection effect. All these stars were chosen in
part based on their X-ray emission (detection by ROSAT), with the idea
that this would preferentially pick out young objects.  Instead it
seems to have favored the inclusion of active objects in the form of
close binaries that are synchronized. The X-ray emission observed in
these cases is therefore not a sign of youth but rather of rapid
rotation maintained by tidal coupling in short-period binaries. We
note that the outer orbit of the triple system HD~97131 is actually
quite long (133.5~d), but it is the secondary (which is itself a
binary) that is probably responsible for the X-ray emission, since its
orbital period is only 1.76~d. 

This highlights the danger of relying heavily on X-ray emission when
trying to identify young stars. Many of them will prove to be older
synchronized binaries. A similar conclusion was reached recently in a
much larger sample of ROSAT-selected sources by \cite{Torres2002}. 
	
\section{Conclusions}
\label{sec:conclusions}

The study of loose associations of young nearby stars has received
considerable attention in the past few years and is beginning to
provide valuable new glimpses into the history of star formation in
the vicinity of the Sun \citep[see, e.g.,][]{Jayawardhana2001}. In
this paper we report spectroscopic results for members and candidate
members of the TW~Hya association based on multiple observations per
object, which have revealed several binary and multiple systems that
had gone unnoticed in many previous studies based on a single
radial-velocity measurement.  Orbital elements have been derived for 5
of these systems, including the systemic velocities. 

None of the 10 potential new members on our list turn out to be true
members of the association, based on our radial velocity results and
measurements of the \ion{Li}{1}~$\lambda$6708 line strength and other
velocity measurements reported by \cite{Song2002}. The fact that many
of them are binaries with short orbital periods is understood as a
selection effect stemming from the use of X-ray emission as one of the
criteria to favor the inclusion of young objects. Although X-ray
emission may indeed be a common characteristic of pre-main sequence
stars, it is shared by a variety of other older active systems as
well. 

A number of objects that do belong to the group and that are known or
suspected to be spectroscopic binaries such as TWA-3A, TWA-5A, TWA-12,
and others, still require radial velocity monitoring to determine
their dynamical properties.  The determination of the line-of-sight
motion for these objects and for other potential members that may be
identified in the future is an important complement to the
proper-motion studies of groups such as the TW~Hya association, and to
our understanding of their relation to larger groups of young stars
within a few hundred parsecs of the Sun. 
	
\acknowledgements

Many of the spectroscopic observations for this project were obtained
by P.\ Berlind, J.\ Caruso, M.\ Calkins, R.\ J.\ Davis, and J.\ Zajac.
and we thank R.\ J.\ Davis for also maintaining the CfA echelle
database.  Helmut Abt, the referee, provided a number of comments and
suggestions that were very helpful.  RN wishes to acknowledge
financial support from the Bundesministerium f\"ur Bildung und
Forschung through the Deutsche Zentrum f\"ur Luft- und Raumfahrt e.V.\
(DLR) under grant number 50~OR~0003.  This research has made use of
the SIMBAD database, operated at CDS, Strasbourg, France, and of
NASA's Astrophysics Data System Abstract Service.

\newpage

\clearpage

\clearpage

\clearpage

\clearpage

\clearpage

\pagestyle{empty}

\begin{deluxetable}{rllccccc}
\tablewidth{40pc}
\tablenum{1}
\tabletypesize{\small}
\tablecaption{Summary of optical properties.\label{tab:sample}}
\tablehead{
\colhead{} & \colhead{} & \colhead{} & \colhead{R.A.} & \colhead{Dec.} &
\colhead{V} & \colhead{} & \colhead{\ion{Li}{1}~$\lambda$6708} \\
\colhead{} & \colhead{TWA \#} & \colhead{Other name} & \colhead{(J2000)} &
\colhead{(J2000)} & \colhead{(mag)} & \colhead{SpT} & \colhead{eq. width (\AA)\tablenotemark{a}}
}
\startdata
 1 &          &  HIP 48273               &  09:50:30.1  &  $+$04:20:37  &  \phn6.24      &  F6            &  0.021$+$0.023     \\
 2 &          &  TYC 6604-0118-1         &  09:59:08.4  &  $-$22:39:35  &  10.09         &  K2            &  0.118             \\
 3 &          &  HIP 50796               &  10:22:18.0  &  $-$10:32:15  &  10.80         &  \nodata       &  $< 0.010$         \\
 4 &          &  HIP 53486               &  10:56:31.0  &  $+$07:23:19  &  \phn7.37      &  K0            &  $< 0.010$         \\
 5 &          &  RXJ1100.0$-$3813        &  11:00:02.4  &  $-$38:13:20  &  12.29         &  \nodata       &  \nodata           \\
 6 &          &  HD~95490                &  11:00:51.2  &  $-$35:33:38  &  \phn8.79      &  F7-8          &  0.13              \\
 7 &  TWA-1   &  TW Hya                  &  11:01:52.0  &  $-$34:42:16  &  \phm{\tablenotemark{b}}10.92\tablenotemark{b} &  K7e            &  0.426             \\
 8 &  TWA-2A  &  CD$-29\arcdeg$8887A     &  11:09:13.9  &  $-$30:01:39  &  11.07         &  M2e           &  0.494             \\
\enddata
\tablenotetext{a}{Average of measurements taken from de la Reza et
al.\ (1989), Favata et al.\ (1995), Soderblom et al.\ (1996), Kastner
et al.\ (1997), Hoff, Henning \& Pfau (1998), Sterzik et al.\ (1999),
Webb et al.\ (1999), Hoff (1999), Torres et al.\ (2000), Zuckerman et
al.\ (2001), Song, Bessell \& Zuckerman (2002), and new measurements
obtained here (see \S2). In the case of the binaries these values are
not corrected for the presence of the companion.}
\tablenotetext{b}{Variable.}
\end{deluxetable}

\clearpage

\begin{deluxetable}{cllcccccccc}
\rotate
\tablewidth{55pc}
\tablenum{2}
\tabletypesize{\small}
\tablecaption{Summary of X-ray properties based on ROSAT observations.\label{tab:xray}}
\tablehead{
\colhead{} & \colhead{} & \colhead{} & \colhead{R.A.\tablenotemark{a}} & \colhead{Dec.\tablenotemark{a}} & \colhead{$\sigma$\tablenotemark{b}} & 
\colhead{} & \colhead{Exposure} & \colhead{Count rate} & \multicolumn{2}{c}{Hardness ratios\tablenotemark{d}} \\
\colhead{} & \colhead{TWA \#} & \colhead{Other name} & \colhead{(J2000)} & \colhead{(J2000)} & \colhead{($\arcsec$)} & 
\colhead{$ML$\tablenotemark{c}} & \colhead{(s)} & \colhead{(s$^{-1}$)} & \colhead{HR1} & \colhead{HR2}
}
\startdata
 1    &         & HIP 48273          &  09:50:30.2  &  $+$04:20:47  & \phn8  &      166  &    417  &  $0.181 \pm 0.022$  &  $-0.15 \pm 0.12$  &  $+0.07 \pm 0.18$ \\
 2    &         & TYC 6604-0118-1    &  09:59:08.2  &  $-$22:39:34  & \phn8  &      230  &    512  &  $0.418 \pm 0.044$  &  $-0.15 \pm 0.10$  &  $-0.12 \pm 0.15$ \\
 3    &         & HIP 50796          &  10:22:19.2  &  $-$10:33:02  &    31  &   \phn18  &    404  &  $0.068 \pm 0.016$  &  $+0.06 \pm 0.26$  &  $-0.05 \pm 0.38$ \\ 
 4    &         & HIP 53486          &  10:56:30.6  &  $+$07:23:14  &    11  &      183  &    344  &  $0.153 \pm 0.022$  &  $-0.36 \pm 0.14$  &  $-0.44 \pm 0.26$ \\
 5    &         & RXJ1100.0$-$3813   &  11:00:05.2  &  $-$38:13:12  &    32  &\phm{22}9  &    244  &  $0.036 \pm 0.015$  &  $+0.42 \pm 0.45$  &  $-0.54 \pm 0.73$ \\
 6    &         & HD~95490           &  11:00:52.8  &  $-$35:33:14  &    24  &   \phn11  &    375  &  $0.045 \pm 0.015$  &  $-0.56 \pm 0.25$  &  $+0.04 \pm 0.67$ \\
 7    & TWA-1   & TW Hya             &  11:01:52.0  &  $-$34:42:12  & \phn7  &      588  &    337  &  $0.571 \pm 0.044$  &  $+0.58 \pm 0.06$  &  $-0.12 \pm 0.08$ \\  
 8    & TWA-2   & CD$-29\arcdeg$8887 &  11:09:13.5  &  $-$30:01:33  & \phn8  &      228  &    325  &  $0.341 \pm 0.035$  &  $-0.22 \pm 0.09$  &  $-0.02 \pm 0.15$ \\
\enddata
\tablenotetext{a}{X-ray position from ROSAT.}
\tablenotetext{b}{Uncertainty in the X-ray position.}
\tablenotetext{c}{Maximum Likelihood estimator (see text).}
\tablenotetext{d}{X-ray hardness ratios are constructed from the
integrated counts in three different ROSAT PSPC energy bands:
0.1--0.4~keV (soft), 0.5--0.9~keV (hard~1), and 0.9--2.0~keV (hard~2).
For the definition of the hardness ratios in terms of the count levels
in these three energy bands, see Neuh\"auser et al.\ (1995).}
\end{deluxetable}

\clearpage

\begin{deluxetable}{rllcccccc}
\rotate
\tablewidth{55pc}
\tablenum{3}
\tabletypesize{\small}
\tablecaption{Effective temperatures and projected rotational velocities for the stars in our sample.\label{tab:teffrot}}
\tablehead{
\colhead{} & \colhead{} & \colhead{} & 
\colhead{} & \colhead{T$_{\rm eff}$ from CfA\tablenotemark{a}} & \colhead{T$_{\rm eff}$ from SpT\tablenotemark{b}} &
\colhead{Other T$_{\rm eff}$\tablenotemark{c}} & \colhead{$v \sin i$ from CfA} &
\colhead{Other $v \sin i$\tablenotemark{c}} \\
\colhead{} & \colhead{TWA \#} & \colhead{Other names} & 
\colhead{SpT} & \colhead{(K)} & 
\colhead{(K)} & \colhead{(K)} & \colhead{($\kms$)} & \colhead{($\kms$)}
}
\startdata
 1 &          &  HIP 48273           &  F6            &  6300$+$6150       &  6530  &         & 22$+$22          & 9$+$14     \\
 2 &          &  TYC 6604-0118-1     &  K2            &  5050$+$4850       &  4840  &         & 19$+$15          & 15, 19, 20 \\
 3 &          &  HIP 50796           &  \nodata       &  4700              &\nodata &         & 2                & 8          \\
 4 &          &  HIP 53486           &  K0            &  5050              &  5150  &         & \phm{:}1:        & 1          \\
 5 &          &  RXJ1100.0$-$3813    &  \nodata       &  5800              &\nodata &         & 34               &            \\
 6 &          &  HD~95490            &  F7-8          &  6450              &  6320  &         & 13               &            \\
 7 &  TWA-1   &  TW Hya              &  K7e           &  4150              &  4150  &  4000, 4150   & 4          & 4, 5, 10, 13, 14, 15 \\
 8 &  TWA-2A  &  CD$-29\arcdeg$8887A &  M2e           &  \phm{:}3670:      &  3520  &  3600   & 13:              & 13, 15     \\
\enddata
\tablenotetext{a}{Values for the coolest objects are based on the
spectral type of the star used as the template, along with the
calibration by de~Jager \& Nieuwenhuijzen (1987).}
\tablenotetext{b}{The calibration adopted is that of de~Jager \&
Nieuwenhuijzen (1987).}
\tablenotetext{c}{Determinations by Fleming (1988), Soderblom et al.\
(1996), Leggett et al.\ (1996), Soderblom et al.\ (1998), Sterzik et
al.\ (1999), Torres et al.\ (2000), Muzerolle et al.\ (2000), Song et
al.\ (2002), and Alencar \& Batalha (2002).}
\end{deluxetable}

\clearpage

\begin{deluxetable}{crrrrc}
\tablenum{4}
\tablewidth{26pc}
\tabletypesize{\scriptsize}
\tablecaption{Radial velocity measurements of HIP~48273 (converted to
the heliocentric frame) and residuals from the orbital
solution.\label{tab:hip48273rvs}}
\tablehead{
\colhead{HJD} & \colhead{RV$_{\rm A}$} & \colhead{RV$_{\rm B}$} &
\colhead{(O$-$C)$_{\rm A}$}  & \colhead{(O$-$C)$_{\rm B}$} & \colhead{Orbital} \\
\colhead{\hbox{~~(2,400,000$+$)~~}} & \colhead{(\kms)} &
\colhead{(\kms)} & \colhead{(\kms)} & \colhead{(\kms)} & \colhead{Phase\tablenotemark{a}} }
\startdata
    45817.6815\dotfill &   $-$31.19 &  $+$68.12  &   $-$0.82  &  $+$2.35  &   0.332 \\
    45817.6942\dotfill &   $-$33.15 &  $+$69.30  &   $-$0.65  &  $+$1.28  &   0.336 \\
    46438.7980\dotfill &   $-$29.50 &  $+$64.58  &   $-$0.50  &  $+$0.26  &   0.670 \\
    46459.7471\dotfill &   $-$76.48 & $+$115.50  &   $+$0.22  &  $+$0.71  &   0.529 \\
    46485.7095\dotfill &  $+$109.15 &  $-$83.38  &   $-$0.29  &  $-$1.21  &   0.028 \\
    46744.9328\dotfill &   $+$89.23 &  $-$60.77  &   $-$0.50  &  $+$0.54  &   0.891 \\
    46771.8911\dotfill &    $-$2.91 &  $+$36.72  &   $+$0.30  &  $-$0.31  &   0.717 \\
    46803.8013\dotfill &   $+$64.73 &  $-$34.95  &   $-$0.53  &  $+$0.47  &   0.163 \\
\enddata
\tablenotetext{a}{Referred to the time of maximum primary velocity, $T_{\rm max}$.}
\end{deluxetable}

\clearpage

\begin{deluxetable}{lccc@{}cc}
\tablenum{5}
\tablecolumns{2}
\tablewidth{27pc}
\tablecaption{Spectroscopic orbital solution for HIP~48273.\label{tab:hip48273elem}}
\tablehead{
\colhead{\hfil~~~~~~~~~~~~~~Parameter~~~~~~~~~~~~~~~} & \colhead{Value}}
\startdata
\multicolumn{2}{l}{Adjusted quantities} \\
~~~~$P$ (days)\dotfill                       &  3.05459836~$\pm$~0.00000084     \\
~~~~$\gamma$ (\kms)\dotfill                  &  $+16.342$~$\pm$~0.066\phs\phn       \\
~~~~$K_{\rm A}$ (\kms)\dotfill               &  94.57~$\pm$~0.12\phn         \\
~~~~$K_{\rm B}$ (\kms)\dotfill               &  100.07~$\pm$~0.17\phn\phn           \\
~~~~$e$\dotfill                              &   0 (fixed)           \\
~~~~$T_{\rm max}$ (HJD$-$2,400,000)\tablenotemark{a}\dotfill    &  48419.18450~$\pm$~0.00045\phm{2222} \\
\multicolumn{2}{l}{Derived quantities} \\
~~~~$M_{\rm A}\sin^3 i$ (M$_{\sun}$)\dotfill &  1.1998~$\pm$~0.0044         \\ 
~~~~$M_{\rm B}\sin^3 i$ (M$_{\sun}$)\dotfill &  1.1338~$\pm$~0.0035       \\
~~~~$q\equiv M_{\rm B}/M_{\rm A}$\dotfill    &  0.9450~$\pm$~0.0020       \\
~~~~$a_{\rm A}\sin i$ (10$^6$ km)\dotfill    &  3.9722~$\pm$~0.0051    \\
~~~~$a_{\rm B}\sin i$ (10$^6$ km)\dotfill    &  4.2032~$\pm$~0.0072      \\
~~~~$a \sin i$ (R$_{\sun}$)\dotfill          &  11.746~$\pm$~0.013\phn      \\
\multicolumn{2}{l}{Other quantities pertaining to the fit} \\
~~~~$N_{\rm obs}$\dotfill                    &  112                      \\
~~~~Time span (days)\dotfill                 &  5816                  \\
~~~~$\sigma_{\rm A}$ (\kms)\dotfill          &  0.85      \\
~~~~$\sigma_{\rm B}$ (\kms)\dotfill          &  1.21      \\
\enddata
\tablenotetext{a}{Time of maximum primary velocity.}
\end{deluxetable}

\clearpage

\begin{deluxetable}{crrrrc}
\tablenum{6}
\tablewidth{26pc}
\tabletypesize{\scriptsize}
\tablecaption{Radial velocity measurements of TYC~6604-0118-1 (converted to
the heliocentric frame) and residuals from the orbital
solution.\label{tab:1e09568rvs}}
\tablehead{
\colhead{HJD} & \colhead{RV$_{\rm A}$} & \colhead{RV$_{\rm B}$} &
\colhead{(O$-$C)$_{\rm A}$}  & \colhead{(O$-$C)$_{\rm B}$} & \colhead{Orbital} \\
\colhead{\hbox{~~(2,400,000$+$)~~}} & \colhead{(\kms)} &
\colhead{(\kms)} & \colhead{(\kms)} & \colhead{(\kms)} & \colhead{Phase\tablenotemark{a}} }
\startdata
    47554.7720\dotfill &   $-$7.62 &  $+$63.36 &    $-$1.33 &    $-$1.15  &   0.670 \\
    47569.8024\dotfill &  $+$64.02 &  $-$20.64 &    $-$1.74 &    $-$4.35  &   0.845 \\
    47575.7061\dotfill &  $+$92.66 &  $-$46.42 &    $+$0.67 &    $-$0.72  &   0.056 \\
    47587.6664\dotfill &  $-$38.88 &  $+$99.61 &    $-$1.84 &    $+$0.62  &   0.561 \\
    47612.6009\dotfill &  $+$76.99 &  $-$28.94 &    $+$0.22 &    $-$0.31  &   0.122 \\
    47640.5398\dotfill &   $+$0.12 &  $+$55.52 &    $+$1.61 &    $-$3.61  &   0.318 \\
    47642.5343\dotfill &  $-$25.94 &  $+$94.36 &    $+$3.53 &    $+$3.86  &   0.403 \\
    47643.5696\dotfill &  $+$92.82 &  $-$48.33 &    $-$1.74 &    $+$0.26  &   0.966 \\
\enddata
\tablenotetext{a}{Referred to the time of maximum primary velocity, $T_{\rm max}$.}
\end{deluxetable}

\clearpage

\begin{deluxetable}{lccc@{}cc}
\tablenum{7}
\tablecolumns{2}
\tablewidth{26pc}
\tablecaption{Spectroscopic orbital solution for TYC~6604-0118-1.\label{tab:1e09568elem}}
\tablehead{
\colhead{\hfil~~~~~~~~~~~~~~Parameter~~~~~~~~~~~~~~~} & \colhead{Value}}
\startdata
\multicolumn{2}{l}{Adjusted quantities} \\
~~~~$P$ (days)\dotfill                       &  1.8386109~$\pm$~0.0000065     \\
~~~~$\gamma$ (\kms)\dotfill                  &  $+27.08$~$\pm$~0.24\phs\phn       \\
~~~~$K_{\rm A}$ (\kms)\dotfill               &  69.07~$\pm$~0.45\phn         \\
~~~~$K_{\rm B}$ (\kms)\dotfill               &  77.46~$\pm$~0.66\phn           \\
~~~~$e$\dotfill                              &   0 (fixed)           \\
~~~~$T_{\rm max}$ (HJD$-$2,400,000)\tablenotemark{a}\dotfill    &  47930.4559~$\pm$~0.0013\phm{2222} \\
\multicolumn{2}{l}{Derived quantities} \\
~~~~$M_{\rm A}\sin^3 i$ (M$_{\sun}$)\dotfill &  0.3168~$\pm$~0.0061         \\ 
~~~~$M_{\rm B}\sin^3 i$ (M$_{\sun}$)\dotfill &  0.2825~$\pm$~0.0046       \\
~~~~$q\equiv M_{\rm B}/M_{\rm A}$\dotfill    &  0.8917~$\pm$~0.0099       \\
~~~~$a_{\rm A}\sin i$ (10$^6$ km)\dotfill    &  1.746~$\pm$~0.012    \\
~~~~$a_{\rm B}\sin i$ (10$^6$ km)\dotfill    &  1.958~$\pm$~0.017      \\
~~~~$a \sin i$ (R$_{\sun}$)\dotfill          &  5.323~$\pm$~0.030      \\
\multicolumn{2}{l}{Other quantities pertaining to the fit} \\
~~~~$N_{\rm obs}$\dotfill                    &  34                      \\
~~~~Time span (days)\dotfill                 &  1157                  \\
~~~~$\sigma_{\rm A}$ (\kms)\dotfill          &  1.68      \\
~~~~$\sigma_{\rm B}$ (\kms)\dotfill          &  2.49      \\
\enddata
\tablenotetext{a}{Time of maximum primary velocity.}
\end{deluxetable}

\clearpage

\begin{deluxetable}{crccc}
\tablenum{8}
\tablewidth{26pc}
\tabletypesize{\small}
\tablecaption{Radial velocity measurements of RXJ1100.0$-$3813 (converted to
the heliocentric frame) and residuals from the orbital
solution.\label{tab:stts-11rvs}}
\tablehead{
\colhead{HJD} & \colhead{RV} & \colhead{$\epsilon$\tablenotemark{a}} &
\colhead{(O$-$C)}  & \colhead{Orbital} \\
\colhead{\hbox{~~(2,400,000$+$)~~}} & \colhead{(\kms)} & \colhead{(\kms)} &
\colhead{(\kms)} & \colhead{Phase\tablenotemark{b}} }
\startdata
    50800.0147\dotfill  &  $+$15.29  &     1.38  &  $+$2.17  &   0.811 \\
    50820.9981\dotfill  &  $+$30.55  &     1.01  &  $+$2.98  &   0.090 \\
    50916.7760\dotfill  &  $+$12.34  &     0.98  &  $-$4.81  &   0.834 \\
    51177.0529\dotfill  &  $-$19.57  &     1.69  &  $-$1.32  &   0.362 \\
    51185.0599\dotfill  &  $+$11.81  &     1.61  &  $-$0.64  &   0.193 \\
    51233.8754\dotfill  &   $+$2.02  &     2.27  &  $+$2.36  &   0.739 \\
    51238.8837\dotfill  &  $-$15.13  &     2.69  &  $+$6.40  &   0.386 \\
    51273.7667\dotfill  &   $+$9.85  &     1.62  &  $+$0.99  &   0.788 \\
\enddata
\tablenotetext{a}{Internal error.}
\tablenotetext{b}{Referred to the time of maximum primary velocity, $T_{\rm max}$.}
\end{deluxetable}

\clearpage

\begin{deluxetable}{lccc@{}cc}
\tablenum{9}
\tablecolumns{2}
\tablewidth{25pc}
\tablecaption{Spectroscopic orbital solution for RXJ1100.0$-$3813.\label{tab:stts-11elem}}
\tablehead{
\colhead{\hfil~~~~~~~~~~~~~~Parameter~~~~~~~~~~~~~~~} & \colhead{Value}}
\startdata
\multicolumn{2}{l}{Adjusted quantities} \\
~~~~$P$ (days)\dotfill                       &  1.373286~$\pm$~0.000031     \\
~~~~$\gamma$ (\kms)\dotfill                  &  $+1.69$~$\pm$~0.82\phs      \\
~~~~$K_{\rm A}$ (\kms)\dotfill               &  30.7~$\pm$~1.1\phn         \\
~~~~$e$\dotfill                              &   0 (fixed)           \\
~~~~$T_{\rm max}$ (HJD$-$2,400,000)\tablenotemark{a}\dotfill    &  51453.9590~$\pm$~0.0074\phm{2222} \\
\multicolumn{2}{l}{Derived quantities} \\
~~~~$a_{\rm A}\sin i$ (10$^6$ km)\dotfill    &  0.58~$\pm$~0.16    \\
~~~~$f(M)$ (M$_{\sun}$)\dotfill              &  0.0041~$\pm$~0.0034 \\
~~~~$M_B \sin i$ (M$_{\sun}$)\dotfill        &  0.1603 $(M_A+M_B)^{2/3}$  \\
\multicolumn{2}{l}{Other quantities pertaining to the fit} \\
~~~~$N_{\rm obs}$\dotfill                    &  20                      \\
~~~~Time span (days)\dotfill                 &  1209                  \\
~~~~$\sigma_{\rm RV}$ (\kms)\dotfill          &  3.23      \\
\enddata
\tablenotetext{a}{Time of maximum primary velocity.}
\end{deluxetable}

\clearpage

\begin{deluxetable}{crrccccc}
\tablenum{10}
\tablewidth{33pc}
\tabletypesize{\scriptsize}
\tablecaption{Radial velocity measurements of HD~97131 (converted to
the heliocentric frame) and residuals from the orbital
solution.\label{tab:stts-8rvs}}
\tablehead{
\colhead{HJD} & \colhead{RV$_{\rm A}$} & \colhead{RV$_{\rm Ba}$} &
\colhead{(O$-$C)$_{\rm A}$}  & \colhead{(O$-$C)$_{\rm Ba}$} & \colhead{Orbital} &
\colhead{Orbital} & \colhead{$\Delta t$\tablenotemark{c}} \\
\colhead{\hbox{~~(2,400,000$+$)~~}} & \colhead{(\kms)} &
\colhead{(\kms)} & \colhead{(\kms)} & \colhead{(\kms)} &
\colhead{Phase\tablenotemark{a}} & \colhead{Phase\tablenotemark{b}} &
\colhead{(days)} }
\startdata 
    50444.0088\dotfill  &  $-$38.08 &   $-$1.69  &  $+$0.69  &  $+$0.69  &  0.422   & 0.819  &   $+$0.0006 \\
    50768.0287\dotfill  &  $-$10.13 &  $-$38.13  &  $-$1.33  &  $-$2.72  &  0.849   & 0.766  &   $+$0.0002 \\
    50772.0430\dotfill  &   $-$9.54 &   $+$2.06  &  $-$1.39  &  $+$3.87  &  0.879   & 0.045  &   $+$0.0000 \\
    50799.0388\dotfill  &  $-$28.57 &  $-$52.77  &  $+$0.16  &  $+$0.40  &  0.081   & 0.371  &   $-$0.0007 \\
    50820.9887\dotfill  &  $-$42.30 &   $+$3.13  &  $-$0.02  &  $+$0.54  &  0.246   & 0.832  &   $-$0.0002 \\
    50821.9875\dotfill  &  $-$42.03 &  $-$49.32  &  $+$0.38  &  $-$0.70  &  0.253   & 0.399  &   $-$0.0001 \\
    50822.9702\dotfill  &  $-$42.59 &  $+$21.17  &  $-$0.08  &  $-$0.32  &  0.261   & 0.957  &   $-$0.0001 \\
    50823.9700\dotfill  &  $-$42.50 &  $-$56.75  &  $+$0.07  &  $-$0.91  &  0.268   & 0.524  &   $-$0.0001 \\
\enddata
\tablenotetext{a}{Referred to the time of periastron passage in the outer orbit.}
\tablenotetext{b}{Referred to the time of maximum primary velocity in the inner orbit.}
\tablenotetext{c}{Correction for light travel time.}
\end{deluxetable}

\clearpage

\begin{deluxetable}{lccc@{}cc}
\tablenum{11}
\tablecolumns{2}
\tablewidth{26pc}
\tablecaption{Spectroscopic orbital solution for HD~97131.\label{tab:stts-8elem}}
\tablehead{
\colhead{\hfil~~~~~~~~~~~~~~Parameter~~~~~~~~~~~~~~~} & \colhead{Value}}
\startdata
\multicolumn{2}{l}{Adjusted quantities in outer orbit (A$-$B)} \\
~~~~$P_{\rm AB}$ (days)\dotfill              &  133.51~$\pm$~0.20\phn\phn     \\
~~~~$\gamma$ (\kms)\dotfill                  &  $-27.09$~$\pm$~0.15\phs\phn      \\
~~~~$K_{\rm A}$ (\kms)\dotfill               &  17.24~$\pm$~0.29\phn         \\
~~~~$K_{\rm B}$ (\kms)\dotfill               &  11.64~$\pm$~0.35\phn         \\
~~~~$e_{\rm AB}$\dotfill                     &  0.191~$\pm$~0.016           \\
~~~~$\omega_{\rm A}$ (deg)\dotfill           &  58.6~$\pm$~4.3\phn           \\
~~~~$T$ (HJD$-$2,400,000)\tablenotemark{a}\dotfill    &  51453.9590~$\pm$~0.0074\phm{2222} \\
\enddata
\tablenotetext{a}{Time of periastron passage.}
\tablenotetext{b}{Time of maximum primary velocity.}
\end{deluxetable}

\clearpage

\begin{deluxetable}{crrrrc}
\tablenum{12}
\tablewidth{30pc}
\tabletypesize{\small}
\tablecaption{Radial velocity measurements of RXJ1115.1$-$3233 (converted to
the heliocentric frame) and residuals from the orbital
solution.\label{tab:stts-10rvs}}
\tablehead{
\colhead{HJD} & \colhead{RV$_{\rm A}$} & \colhead{RV$_{\rm B}$} &
\colhead{(O$-$C)$_{\rm A}$}  & \colhead{(O$-$C)$_{\rm B}$} & \colhead{Orbital} \\
\colhead{\hbox{~~(2,400,000$+$)~~}} & \colhead{(\kms)} &
\colhead{(\kms)} & \colhead{(\kms)} & \colhead{(\kms)} & \colhead{Phase\tablenotemark{a}} }
\startdata
    50800.0462\dotfill  &  $+$33.56 &  $-$61.33  &   $+$1.50  &   $-$1.74  &   0.824 \\
    50821.0455\dotfill  &   $+$2.49 &   $+$9.75  &   $+$2.59  &  $+$26.05  &   0.237 \\
    50884.8566\dotfill  &  $+$38.29 &  $-$85.97  &   $-$1.55  &  $-$15.91  &   0.841 \\
    50916.7556\dotfill  &  $+$47.34 &  $-$89.46  &   $-$1.13  &   $-$7.77  &   0.140 \\
    51177.0286\dotfill  &  $+$25.40 &  $-$57.81  &   $+$1.74  &   $-$9.52  &   0.808 \\
    51185.0241\dotfill  &  $-$73.92 &  $+$85.24  &   $+$0.44  &   $+$1.55  &   0.392 \\
    51187.0223\dotfill  &  $-$29.69 &  $+$20.21  &   $-$2.59  &   $+$0.16  &   0.287 \\
    51233.8874\dotfill  &  $-$29.20 &  $+$17.92  &   $+$1.79  &   $-$7.37  &   0.295 \\
    51235.8818\dotfill  &  $+$25.68 &  $-$47.76  &   $+$0.03  &   $+$3.20  &   0.189 \\
    51238.8703\dotfill  &  $-$91.80 &  $+$98.68  &   $+$0.59  &   $-$9.28  &   0.528 \\
    51272.7801\dotfill  &  $-$19.62 &   $-$9.80  &   $-$0.84  &  $-$18.66  &   0.728 \\
    51273.7814\dotfill  &  $+$31.16 &  $-$63.32  &   $-$0.17  &   $-$4.71  &   0.177 \\
    51292.7703\dotfill  &  $-$40.55 &  $+$30.12  &   $-$1.09  &   $-$6.57  &   0.689 \\
    51325.7099\dotfill  &  $-$89.40 & $+$110.93  &   $+$0.79  &   $+$5.94  &   0.454 \\
    51540.0251\dotfill  &  $-$94.15 & $+$101.37  &   $-$1.19  &   $-$7.36  &   0.521 \\
\enddata
\tablenotetext{a}{Referred to the time of maximum primary velocity, $T_{\rm max}$.}
\end{deluxetable}

\clearpage

\begin{deluxetable}{lccc@{}cc}
\tablenum{13}
\tablecolumns{2}
\tablewidth{25pc}
\tablecaption{Spectroscopic orbital solution for RXJ1115.1$-$3233.\label{tab:stts-10elem}}
\tablehead{
\colhead{\hfil~~~~~~~~~~~~~~Parameter~~~~~~~~~~~~~~~} & \colhead{Value}}
\startdata
\multicolumn{2}{l}{Adjusted quantities} \\
~~~~$P$ (days)\dotfill                       &  2.230887~$\pm$~0.000024     \\
~~~~$\gamma$ (\kms)\dotfill                  &  $-7.00$~$\pm$~0.41\phs       \\
~~~~$K_{\rm A}$ (\kms)\dotfill               &  86.75~$\pm$~0.71\phn         \\
~~~~$K_{\rm B}$ (\kms)\dotfill               &  116.8~$\pm$~5.2\phn\phn           \\
~~~~$e$\dotfill                              &   0 (fixed)           \\
~~~~$T_{\rm max}$ (HJD$-$2,400,000)\tablenotemark{a}\dotfill    &  51137.3021~$\pm$~0.0020\phm{2222} \\
\multicolumn{2}{l}{Derived quantities} \\
~~~~$M_{\rm A}\sin^3 i$ (M$_{\sun}$)\dotfill &  1.12~$\pm$~0.12         \\ 
~~~~$M_{\rm B}\sin^3 i$ (M$_{\sun}$)\dotfill &  0.831~$\pm$~0.049       \\
~~~~$q\equiv M_{\rm B}/M_{\rm A}$\dotfill    &  0.743~$\pm$~0.037       \\
~~~~$a_{\rm A}\sin i$ (10$^6$ km)\dotfill    &  2.661~$\pm$~0.024    \\
~~~~$a_{\rm B}\sin i$ (10$^6$ km)\dotfill    &  3.58~$\pm$~0.18      \\
~~~~$a \sin i$ (R$_{\sun}$)\dotfill          &  8.97~$\pm$~0.25      \\
\multicolumn{2}{l}{Other quantities pertaining to the fit} \\
~~~~$N_{\rm obs}$\dotfill                    &  15                      \\
~~~~Time span (days)\dotfill                 &  740                  \\
~~~~$\sigma_{\rm A}$ (\kms)\dotfill          &  1.55      \\
~~~~$\sigma_{\rm B}$ (\kms)\dotfill          &  11.8      \\
\enddata
\tablenotetext{a}{Time of maximum primary velocity.}
\end{deluxetable}

\clearpage

\begin{deluxetable}{crc}
\tablenum{14}
\tablewidth{19pc}
\tabletypesize{\small}
\tablecaption{Radial velocity measurements (converted to
the heliocentric frame) for stars without orbital solutions.\label{tab:tabrvs}}
\tablehead{
\colhead{HJD} & \colhead{RV} & \colhead{$\epsilon$\tablenotemark{a}} \\
\colhead{\hbox{~~~~~(2,400,000$+$)~~~~~}} & \colhead{(\kms)} & \colhead{(\kms)}}
\startdata
\multicolumn{3}{l}{HIP~50796} \\
  ~~~~46422.9750\dotfill  &   $+$16.31  &   0.22 \\
  ~~~~46537.6905\dotfill  &   $+$14.06  &   0.41 \\
  ~~~~46540.7560\dotfill  &   $+$12.68  &   0.29 \\
  ~~~~46569.7043\dotfill  &   $+$12.02  &   0.52 \\
  ~~~~52391.6615\dotfill  &   $+$47.98  &   0.60 \\
  ~~~~52395.7197\dotfill  &   $+$48.09  &   0.42 \\
  ~~~~52419.6668\dotfill  &   $+$40.40  &   0.49 \\
\enddata
\tablenotetext{a}{Internal error.}
\tablenotetext{b}{Measurement from Tautenburg.}
\tablenotetext{c}{Measurement from La Silla (FEROS).}
\end{deluxetable}

\clearpage

\begin{deluxetable}{rllcc@{}cl}
\tablewidth{41pc}
\rotate
\tablenum{15}
\tabletypesize{\scriptsize}
\tablecaption{Summary of radial velocity results for our sample.\label{tab:meanrvs}}
\tablehead{
\colhead{} & \colhead{} & \colhead{} & \colhead{} &
\colhead{Mean CfA RV} & \colhead{Other RVs\tablenotemark{a}} & \colhead{} \\
\colhead{} & \colhead{TWA \#} & \colhead{Other names} & \colhead{N$_{\rm obs}$} & 
\colhead{($\kms$)} & \colhead{($\kms$)} & \colhead{Remarks}
}
\startdata
 1 &          &  HIP~48273              & 112       &  $+$16.342~$\pm$~0.066\phm{$+1$}     & $+$17.0, $+$17.4            & Double-lined binary (orbit) \\
 2 &          &  TYC~6604-0118-1        & 34        &  $+$27.08~$\pm$~0.24\phm{$+1$}       & $-$19, $+$54                & Double-lined binary (orbit) \\
 3 &          &  HIP~50796              & 10        &  \nodata                             & $+$22.4                     & Variable RV \\
 4 &          &  HIP~53486              & 7         &  $+$5.47~$\pm$~0.26\phm{$+$}         & $+$4.3                      & \\
 5 &          &  RXJ1100.0$-$3813       & 20        &  $+$1.69~$\pm$~0.82\phm{$+$}         &                             & Single-lined binary (orbit) \\
 6 &          &  HD~95490               & 4         &  $-$7.51~$\pm$~0.17\phm{$-$}         &                             & \\
 7 &  TWA-1   &  TW~Hya                 & 8         &  $+$12.66~$\pm$~0.22\phm{$+1$}       & $+$12, $+$12.5, $+$13.5, $+$13 & \\
 8 &  TWA-2A  &  CD$-$29$\arcdeg$8887A  & 9         &  $+$10.97~$\pm$~0.12\phm{$+1$}       & $+9.7$, $+10$, $+12$           & \\
\enddata
\tablenotetext{a}{Measurements by Popper \& Shajn (1948), Mayor \&
Mazeh (1987), de la Reza et al.\ (1989), Stauffer et al.\ (1995),
Reipurth, Pedrosa \& Lago (1996), Sterzik et al.\ (1999), Torres et
al.\ (2000), Barbier-Brossat \& Figon (2000), Neuh\"auser et
al.(2000b), and Song et al.\ (2002).}
\end{deluxetable}

\clearpage

\begin{deluxetable}{cllccccc}
\tablewidth{41pc}
\tablenum{16}

\tablecaption{Members and candidate members from the kinematic study by Makarov
\& Fabricius (2001) with measured radial velocities.\label{tab:makarov}}

\tablehead{
\colhead{} & \colhead{} & \colhead{} & \colhead{$\mu$}  & \colhead{$D_{\rm kin}$} & \colhead{$D_{\rm trig}$} & \colhead{$RV_{\rm pred}$} & \colhead{$RV_{\rm obs}$} \\
\colhead{} & \colhead{TWA \#} & \colhead{Other names} & \colhead{(mas~yr$^{-1}$)} & \colhead{(pc)} & \colhead{(pc)} & \colhead{(km~s$^{-1}$)} & \colhead{(km~s$^{-1}$)}
}
\startdata
\multicolumn{8}{l}{Objects with velocities reported in this paper} \\
 1    &           &  HIP~48273              & 157.9                & 26.6      &  45.9     & $+$10.7           &  $+$16.3\phm{$+$}    \\
 2    &           &  TYC~6604-0118-1        & \phn63.7             & 62.8      & \nodata   & $+$16.5           &  $+$27.1\phm{$+$}    \\
 3    &           &  HIP~50796              & \phn79.1             & 53.8      &  34.0     & $+$13.1           &  \nodata             \\
 4    &           &  HIP~53486              & 268.1                & 16.7      &  17.6     & \phantom{1}$+$3.7 &  \phn$+$5.5\phm{$+$} \\
 7    &  TWA-1    &  TW~Hya                 & \phn75.4             & 57.1      &  56.4     & $+$12.7           &  $+$12.7\phm{$+$}    \\
 8    &  TWA-2A   &  CD$-$29$\arcdeg$8887A  & \phn92.6             & 47.1      & \nodata   & $+$10.6           &  $+$11.0\phm{$+$}    \\
17/18 &  TWA-4A/B &  HD~98800A/B            & \phn96.8             & 45.7      &  46.7     & \phantom{1}$+$9.1 &  \phn$+$9.2\phm{$+$} \\
\enddata
\end{deluxetable}

\clearpage

\begin{deluxetable}{llcccccccccc}
\rotate
\tablewidth{51pc}
\tablenum{17}
\tabletypesize{\tiny}
\tablecaption{Objects currently considered members of the TW~Hya association.\label{tab:alltwa}}
\tablehead{
\colhead{} & \colhead{} & \colhead{R.A.\tablenotemark{a}} & \colhead{Dec.\tablenotemark{a}} & \colhead{} & 
\colhead{$V$\tablenotemark{b}} & \colhead{$B\!-V$} & \colhead{$\mu_{\alpha}$\tablenotemark{a}} & \colhead{$\mu_{\delta}$\tablenotemark{a}} & \colhead{$\pi_{\rm HIP}$} & 
\colhead{Li~I~$\lambda$6708} & \colhead{Mean RV} \\
\colhead{} & \colhead{Other name} & \colhead{(J2000)} & \colhead{(J2000)} & \colhead{SpT} & 
\colhead{(mag)} & \colhead{(mag)} & \colhead{(mas~yr$^{-1}$)} & \colhead{(mas~yr$^{-1}$)} & \colhead{(mas)} &
\colhead{eq.\ width (\AA)} & \colhead{(km~s$^{-1}$)}
}
\startdata
TWA-1   &  TW Hya                 &  11:01:51.9  &  $-$34:42:17  &  K7e     &  10.92 &  0.97 &   $-$73.4 &  $-$17.5 &  17.72 &  0.426         &  $+$12.66~$\pm$~0.22\phm{$+1$} \\
\noalign{\vskip 5pt}
TWA-2A  &  CD$-29\arcdeg$8887A    &  11:09:13.8  &  $-$30:01:40  &  M2e     &  11.07 &  1.42 &   $-$90.1 &  $-$21.1 &\nodata &  0.494         &  $+$10.97~$\pm$~0.12\phm{$+1$} \\
TWA-2B  &  CD$-29\arcdeg$8887B    &    $+0\farcs3$ &    $+0\farcs5$  &  M2      & \phm{1K}$+0.81$K\phm{$+$} &\nodata &\nodata  &\nodata &\nodata &  \nodata       &  \nodata \\
\noalign{\vskip 5pt}
TWA-3A  &  Hen~3-600A             &  11:10:28.0  &  $-$37:31:52  &  M3e     &  12.04 &  1.52 &   $-$96.5 &  $+$16.3 &\nodata &  0.563         &  $+$11.2 \\
TWA-3B  &  Hen~3-600B             &    $-0\farcs8$  &    $-1\farcs2$   &  M3.5    & \phm{1K}$+0.52$K\phm{$+$} &\nodata&\nodata  &\nodata &\nodata &  0.52          &  $+$14.0 \\
\noalign{\vskip 5pt}
TWA-4A  &  HD 98800A              &  11:22:05.3  &  $-$24:46:40  &  K4-5    &  \phn9.41 &  1.15 &   $-$91.7 &  $-$31.1 &  21.43 &  0.425         &  $+$12.75~$\pm$~0.10\phm{$+1$} \\
TWA-4B  &  HD 98800B              &    $+0\farcs0$   &    $+0\farcs8$  &  K7$+$M1 &  \phn9.94 &  1.28 & \nodata & \nodata&\nodata &  0.335$+$0.450 &  $+$5.73~$\pm$~0.14\phm{$+$} \\
\noalign{\vskip 5pt}
TWA-5A\tablenotemark{c}  &  CD$-33\arcdeg$7795A    &  11:31:55.3  &  $-$34:36:27  &  M1.5    &  11.54 &  1.48 &   $-$81.6 &  $-$29.4 &\nodata &  0.572         &  \nodata \\
\enddata

\tablenotetext{a}{Positions (epoch 2000) and proper motions are taken preferentially from the Tycho-2 catalog (H\o g et al.\ 2000) or from the UCAC1 catalog (Zacharias et al.\ 2000), when available.}
\tablenotetext{b}{Magnitudes or magnitude differences available only in the $R$ or $K$ bands, instead of $V$, are labeled.}
\tablenotetext{c}{Close visual binary with a separation of $0\farcs0548$ and a magnitude difference $\Delta H = 0.09$ (Macintosh et al.\ 2001).}
\tablenotetext{d}{Although formally the magnitude difference indicates
a brighter secondary (Zuckerman et al.\ 2001b), measurement errors in
both the magnitudes and spectral types suggest the stars are
indistinguishable within the errors (Schwartz \& Zuckerman 2002,
priv.\ comm.).}
\tablenotetext{e}{Close visual binary with a separation of $0\farcs67$ and a brightness ratio $\sim$0.9 in $H$ (see Zuckerman et al.\ 2001b).}
\tablenotetext{f}{Proper motions relative to other stars in the field (Gizis 2002).}
\end{deluxetable}

\end{document}